\documentclass[12 pt]{article}
\newtheorem{theorem}{Theorem}
\newtheorem{definition}{Definition}
\begin{document}
\title{Hilbert's 6-th Problem\\ and Axiomatization of Dynamics}
\author{V.~Yu. Tertychny-Dauri\\ Saint-Petersburg State University of\\ Information Technologies, Mechanics and Optics\\ (University ITMO)\\ Department of Physics and Engineering\\ 49, Kronverkskyi Pr., Saint-Petersburg 197101, Russia\\ email: tertychny-dauri@mail.ru}
\date{26 October 2020}
\maketitle
\begin{abstract}
The following offers a new axiomatic basis of mechanics and physics in their most important dynamics domain, i.~e. an axiom (principle) of completeness intended to generalize Newton's second law of motion for the case of a non-stationary variable-mass point (system) that varies with time. This generalization leads to hyperdynamic dependencies describing such motion from new accurate qualitative standpoints.
\end{abstract}
\newpage
\section*{Introduction}
This article can be considered a modified version of {\it the theory of hyperreactive motion} proposed by the author as far back as in 1991 [1, 2]. The incentive was to contribute a mite to the solution of David Hilbert's sixth physical problem [3, 4]. Let us recall that, in this rather generic problem, a not definitely specified task of mathematical axiomatization of mechanics and physics as a whole has been set. 

Being mainly based on the results obtained by I. V.$\,$Meshchersky [5], K. E.$\,$Tsiolkovsky [6, 7], T.$\,$Levi--Civita [8], the theory of reactive motion, in spite of its seemingly completed and elaborated forms, nevertheless gives rise to the two natural (although not equally significant) questions. The first one: if the absolute velocity $u$ of the particle outflow (radiating flow) does not satisfy the condition of stationarity, what should the structure of motion equations be in such case, and what additional forces might arise due to the non-stationarity effect?

The second question is closely related to the first one and is crucial in the context of generic mechanical constructions: why do standard equations of reactive motion contain only the mass of a system (point) $M(t)$ at a given instant $t$ and the speed of its change $dM(t)/dt$, and do not contain, as a legitimate term, the value which includes the acceleration of the object mass change $d^2 M(t)/dt^2$?

The last fact might seem even more strange because since Newton and Lagrange the major focus in dynamics was put on terms of that kind. In the following paragraphs, we will try to answer these and some other questions concerning non-stationary reactivity using the model of hyperreactive motion [1, 2, 9, 10].

There exist ample reasons to think that the so-called $u$-factor (i.e. essential impossibility of practically accurate, direct calculation of the absolute exhaust velocity $u$ of emitted particles) is an unnoticed source for various disagreements in hydro- and aerodynamics, as well as in reactive and space dynamics theories.

The spread in measured values of $u$ can be quite significant and can vary in a rather wide range. In such circumstances, the exact measurement parameter remains the end dynamic result in the form of a certain force load: pressure, thrust force, reactive force, etc. Depending on this final dynamic "product", the exhaust velocity $u$ is calculated --- in this particular order, and not in an inverse one. If the calculation procedure is flawed, then, evidently, the exhaust velocity $u$ will be calculated not quite accurately.

Introduction of the new hyperreactive forces into analysis also required a qualitatively new structural approach to the basic dynamic principles of mechanics, their refinement and modernization. In this sense, the results of Meshchersky--Levi--Civita's theory on derivation of equations of reactive motion, as well as Tsiolkovsky's calculation scheme cannot be deemed entirely satisfactory. There is a vast literature dealing with delicate and complicated questions of the motion of dynamic systems with variable mass (e.~g. [5, 6, 7, 8, 11, 12, 13, 14, 15, 16, 17, 18]); however, any new development of a variable-mass model which results in appearing of additional and efficient forces deserves the closest attention and evaluation.

In Section 1, the concept of a new type of reactive motion, namely, hyperreactive motion of a variable-mass point, is substantiated. The key point of the new approach is consideration of hyperreactive components, i.e. taking into account summands that depend on the acceleration of the point mass change and take part in the dynamic description of the system motion on equal terms with other components. The new axiomatic principle of dynamics was called "the axiom (principle) of completeness".

The next Section 2 introduces the simplest applications of the hyperreactive model. These applications relate to well-known spaceflight dynamics problems of Tsi\-ol\-kov\-sky and an exterior ballistics issue. Hyperreactive equations of motion are derived, and some integral characteristics of motion are obtained.

Then, in Section 3, Tsiolkovsky's problems are extensively studied in hyperreactive formulation. Integrals of motion are determined. Special attention is paid to searching for laws of mass change, wherein the system has a given character of motion. Optimal regimes of motion were studied, and parameters were found that ensure these regimes of the hyperreactive system operation.

In Section 4, new concepts of reactive and efficient energy of a variable-mass point are introduced for the hyperreactive motion, and the theorem of efficient energy change is substantiated. After that, transition to generalized curvilinear coordinates is performed, and hyperreactive Lagrange equations of the second kind are derived in the curvilinear coordinate system. The section ends with the formulation of Hamilton's principle in the hyperreactive case.

In order to write down Hamilton's principle in a traditional form, a variational integral is introduced in Section 5. This class of integrals proves to be so efficient that it allows for performing various transformations containing the operation of variation of functionals (complicated functions). Using the variational integral, it is possible to relatively easily obtain the formulation of Hamilton's principle and Lagrange equations for the hyper-motion in a standard form.

\section{Axioma (principle) of completeness}
This section substantiates of the most important dynamic principle in mechanics of variable-mass point, namely the {\it principle of completeness}, which allows for approaching the problem of non-stationary reactive motion from new standpoints.

A (symmetrical) hyperreactive model of motion of a variable-mass point is proposed below, which allows for accurate consideration of force actions. The hyperreactive model based on the new differential law of motion (principle of completeness) contains those summands in the equation of motion that depend not only on point mass $M(t)$ at a time moment $t$ and on the speed of its change $dM(t)/dt,$ but also on the acceleration of the mass change $d^2M(t)/dt^2,$ which is crucially important from the perspective of global description of material bodies' motion process. 

Consider the motion of a material point, the mass $M(t)$ of which changes continuously and smoothly enough over the time interval $[\,t_0,\,t\,]$. For definiteness, let us assume that $M(t)$ decreases due to fact that the point has a radiating flow of particles moving with the absolute vector velocity $u$ (hereinafter, vector values are not marked to achieve simplicity). If the work of the radiating center is known, it is assumed that $M(t)$ is a known time function.

If the vector function $u(t)$ defined over the interval $[\,t_0,\,t\,]$ is stationary, i.~e. $du(t)/dt = 0,\,u(t) = {\rm const},$ then Meshchersky's equation of motion takes the form
$$M\,{{dv}\over{dt}} = F + {{dM}\over{dt}}\,(u - v),\eqno(1)$$
where $v(t)$ is the absolute velocity of the point, and $F$ is the external active force acting on the point.

Equation (1) is usually written down as follows:
$$M\,{{dv}\over{dt}} = F + \Phi_1,\qquad \Phi_1 = {{dM}\over{dt}}\,V,\eqno(2)$$
where $\Phi_1$ is the reactive force, $V = u - v$ is the relative velocity of emitted particles, $|\,dM/dt\,|$ is the mass flow of the radiating center per unit of time. Furthermore, using the theorem of momentum for the value
$Q_1(t) = M(t)\,v(t),$ we obtain the following from equations (1), (2):
$${{dQ_1}\over{dt}} = F + {{dM}\over{dt}}\,u.$$

Let us recall that the derivation of equations (1), (2) is based on not quite correct {\it Meshchersky's hypothesis of close-range interaction}, which allows for considering only emitted particles with the mass $dM$ and using only the value of momentum of the point in the form $Q_1 = Mv.$

A new model of interaction of the system "point--particle" will be proposed below which can be called the model of equally significant symmetrical interaction. In this model, the point and the particle have momenta of the same order and play quasi the same role --- the role of a fully legitimate participant of the interaction. Essentially, it makes no difference here, what separates from what: the point from the particle, or the particle from the point; the difference between them is conditional.

One can therefore speak about their {\it relative velocity} $V = u - v$ of motion of the point with respect to the particle (flow of particles); and equally reasonably one can speak about the relative velocity of motion of the particle with respect to the point. The {\it symmetrical model of interaction} of the remaining and separated mass takes into account the influence of the latter on the dynamics of the remaining part, and vice versa.

In our analysis, let us proceed from the following important axiom: the concepts of momentum or measure of motion are of universal, i.e. absolute, nature [9, 10] with regard to systems with changing parameters (including varying masses). These concepts should include all velocity characteristics of the dynamical state of the considered system $(v,\,u, \,dM/dt).$

As the object dynamics is described in generalized independent coordinates, one more independent variable, i.e. mass, should be added to the space coordinates as a new basis coordinate. The effect of change of mass on the dynamics of the entire system is determined using the equation of motion. This approach underlies the principle of completeness.

Let us proceed with the progressive realization of the program outlined above.
We start with the simplest case when it is assumed that the exhaust velocity of particles $u(t)$ is non-stationary $u(t) \not= {\rm const},\ \forall\,t \in [\,t_0,\,t
\,].$ As the particle acts upon the material point with mass $M(t)$ at a time moment $t$ with this velocity (more correctly, with velocity $- u)$, it is reasonable to consider, apart from momentum $Q_1(t)$, an additional momentum $Q_2(t) = M(t)\,u(t)$, and the corresponding force 
$${{dQ_2}\over{dt}} = {{dM}\over{dt}}\,u + M\,{{du}\over{dt}}.\eqno(3)$$

In this case, the point moves with relative velocity $v - u$ and experiences the resulting momentum $Q_* = Q_1 - Q_2 = M\,(v - u) = -\,MV;$ the equation of motion itself, taking into account expression (3), can be written down as follows:
$${{dQ_1}\over{dt}} = F + {{dM}\over{dt}}\,u + M\,{{du}\over{dt}},$$
and hence
$$M\,{{dv}\over{dt}} = F + {{dM}\over{dt}}\,(u - v) + M\,{{du}\over{dt}}.\eqno(4)$$

Equation (4) in terms of its form is a modified variant of Meshchersky's equation; the third term in its right-hand side is the additional reactive force $\Phi_2$ that arises due to the {\it non-stationarity effect}:
$$M\,{{dv}\over{dt}} = F + \Phi_1 + \Phi_2,$$
or
$$F + {{(MV)}\over{dt}} = 0.$$
In terms of momentum $Q_*$, equation (4) has a simpler and more compact form:
$${{dQ_*}\over{dt}} = F,\qquad Q_* = Q_1 - Q_2,\eqno(5)$$
which corresponds to the principle of momentum.

Let us note that the stationary model of Meshchersky (1) cannot be deemed satisfactory for a large class of problems because of noncompliance of this model with the real behavior of dynamic processes.

Furthermore, we should also notice that the non-stationary model of Meshchersky (5) assumes that the system possesses an accelerating non-stationary radiating center (for example, engine-accelerator) that ensures the appearance of the additional force.

At the same time, it would be premature to assume that equation (5) provides the most complete and correct description of dynamics of a variable-mass point. 
Arguments set forth here are one of the logical steps of the stair to be surmounted before consideration of a more generic model.

Based on the proposed concept of the universal velocity nature of the momentum and the complete set of independent variables (coordinates), one should continue increasing the value of momentum acting upon the point, i.e. it is necessary to include the value depending on the speed of mass change $dM(t)/dt.$

Assuming the mass to be one more generalized coordinate, the {\it total momentum} of the point (system) can thus be written as follows:
$$Q(t) = M(t)\,{{dR(t)}\over{dt}} + {{dM(t)}\over{dt}}\,R(t),\eqno(6)$$
where vector $R(t) = r(t) - \rho(t)$, let us call it {\it reactive vector}, $dr(t)/dt =$ $= v(t)$ is the absolute velocity of motion of the point, $d\rho(t)/dt = u(t)$ is the absolute exhaust velocity of particles of the radiating center that represents a time vector function defined over $[\,t_0,\,
t\,]$. Here $r(t),\,\rho(t)$ are current radius vectors of the point and the particle in the absolute frame of reference, correspondingly.

The second term in the right-hand side of equation (6) represents the momentum that appears due to the change of mass over the space vector segment $R(t).$ If we denote this term with
$$Q_3(t) = {{dM(t)}\over{dt}}\,R(t),\eqno(7)$$
then, taking into account (7), the total momentum of the point will be equal to $Q = Q_1 - Q_2 + Q_3.$

We thus see that the total resulting momentum $Q(t)$ represents a sum of the standard momentum $M(t)\,\bigl(dR(t)/dt\bigr)$ that arises due to the presence of the relative velocity $dR(t)/dt$ of the point (system) with mass $M(t)$ and momentum $\bigl(dM(t)/dt\bigr)\,R(t)$ that arises, as was already mentioned, due to the speed of change of the mass $dM(t)/dt$ along the vector $R(t).$

In accordance with expression (6), the momentum of the system "point -- elementary particle" over the time $dt$ equals to
$$Q(t)\,dt = M(t)\,dR(t) + \bigl(dM(t)\bigr)\,R(t),$$
i.~e. the point mass $M(t)$ is transported for a small distance $dR(t);$ at the same time, it decreases, and the elementary mass $dM(t)$, while being moved to the category of emitted particles, passes the distance $R(t).$ In other words, the mass $M(t)$ during the time $dt$ passes the distance $dr,$ but, as a certain part separates from it during this time, $M(t)$ moves backwards over the distance $d\rho.$ The mass $M(t)$ is actually affected by the resulting relative velocity $dR/dt = v - u.$

Let us also note that the elementary mass $dM(t)$ belongs to the sophisticated dialectic category of masses: on the one hand, it belongs to still remaining mass of the point $M(t),$ but, on the other hand, it becomes something like a part of the separated mass of particles (particle flow). During the time $dt$, the deep process of interaction and redistribution of masses occurs, and the separation phase is formed.

Let us call the product of the point mass $M(t)$ and the reactive vector $R(t)$ the {\it motion composition} vector $S(t) = M(t)\,R(t).$ Then, in accordance with expression (6), we obtain the new differential law of motion: axiom (principle) of completeness or the theorem of changing of the motion composition.

\begin{theorem}
Time derivative of the motion composition vector of a system equals to the vector of its total momentum:
$${{dS(t)}\over{dt}} = Q(t).\eqno(8)$$
\end{theorem}

Concerning the law of motion (8), the following observations can be made.

{\bf Remarks.} 1. The resulting law of motion is universal in the sense that it corresponds to the local laws of dynamics that constitute differential laws of changing (of momentum, angular momentum, and kinetic energy).

2. From the structure of the motion composition, one can readily see that this concept is the closing one in dynamics, and it cannot be subdivided into components with "lower" differential dimensions, for example, $\int_{t_0}^t R(s)\,ds.$ The principle of completeness thus acquires features of the generic and basic dynamic law.

Now let us proceed with the most important part of the analysis: writing down equations of motion of a variable-mass material point. We use the momentum principle for the value $Q(t)$ (6):
$${{dQ(t)}\over{dt}} = F(t),$$
where $F(t)$ is the external active force acting on the point.

We get
$$M\,{{d^2R}\over{dt^2}} + {{dM}\over{dt}}\,{{dR}\over{dt}} + {{dM}\over{dt}}\,
{{dR}\over{dt}} + {{d^2M}\over{dt^2}}\,R = F,$$
thus
$$M\,{{d^2R}\over{dt^2}} + 2\,{{dM}\over{dt}}\,{{dR}\over{dt}} + {{d^2M}\over
{dt^2}}\,R = F$$
and, hence,
$$M\,{{dv}\over{dt}} = F + 2\Phi_1 + \Phi_2 + \Phi_3,\eqno(9)$$
where the following notations are introduced:
$$\Phi_1 = -\,{{dM}\over{dt}}\,{{dR}\over{dt}} = {{dM}\over{dt}}\,V,\quad \Phi_2
= M\,{{du}\over{dt}},\quad \Phi_3 = -\,{{d^2M}\over{dt^2}}\,R.$$
Here $\Phi_1$ is the standard {\it reactive force} with the relative exhaust velocity of particles $V = u - v,\ \Phi_2$ is the force arising due to the effect of non-stationarity of velocity $u.$ The force $\Phi_3$ generated by the acceleration of the point mass change will be referred to as {\it hyperreactive}.

Let us add that, in terms of the reactive vector $R(t)$ and the relative velocity $\dot{R}(t) = - V(t)$, non-stationary Meshchersky's equation of the reactive motion looks as follows:
$${{d\,\bigl(M\,(v - u)\bigr)}\over{dt}} = F,\quad {\hbox{or}}\quad {{d\,(M\dot{
R})}\over{dt}} = F,$$
and the general {\it equation of hyperreactive motion} is written down as
$${{d\,(M\dot{R} + \dot{M}R)}\over{dt}} = F,$$
where the over-dot denotes the time derivative. One can clearly see due to which term and why the effect of hyperreactivity appears.

Let us comment the derived equations of hyperreactive motion (9). The most interesting point concerning equation (9) is the comparison with Meshchersky's equation that can be represented in form (2). As we can see, the new formulation undergoes considerable changes. Even when we deal with the stationary emission of particles, and $d^2 M/dt^2 = 0,$ one cannot speak about similar equations.

The equations of motion could not be the same, because different values were selected as momentum. In Meshchersky's model, it is essentially impossible to consider terms containing $d^2M/dt^2,$ and this is its utmost disadvantage, which cannot be said about the new model. From the formal point of view, one should choose the total resulting momentum $Q$ rather than the standard momentum $Q_1$ as the generic momentum.

The difference in the calculation of the stationary reactive force can be explained as follows (using the $u$--factor): actual, exact determination of the absolute exhaust velocity of particles $u,$ and hence the relative velocity $V,$ is very cumbersome and, as is well known, represents a complicated technical problem; errors during measurement and calculation can be quite significant.

Thus, firstly, the picture of motion, even being represented by "distorted" Meshchersky's model, can inspire a certain confidence, especially in case of stationarity and relatively low velocities of the emitted particles.

Secondly, ignoring terms $\Phi_2$ and $\Phi_3$ in relation (9) at different stages of motion has resulted in noticeable inaccuracies in trajectories of dynamic objects, in excessive correction and energy consumption. Based on empirical data, these terms might be identified as various random and deterministic perturbations.

Let us recall the work On Action of Fluids on Rigid Bodies and on Motion of Rigid Bodies in Fluids by D.$\,$Bernoulli (mentioned in Ch. 1 of book [10]), where the author states that the reaction force of a jet of fluid (reactive force) should have the double coefficient. Other variants, in his opinion, do not comply with the experience because they "would resemble enough the true ones if they better corresponded to experimental values". Unfortunately, his practical conclusions based on thorough measurement of reactive jet speed were not supported by theoretical analysis. Nevertheless, D.$\,$Bernoulli can be deemed a precursor of the hyperreactive mechanics.

In addition to the information on coefficient 2 at the reactive force $\Phi_1$, we would like to note that field tests of rocket jet engines aimed at measurement of force characteristics $F_T$ show just that the relative exhaust velocity $V$ of gases from the nozzle (in the considered case, when $v = 0$, we have $V = u)$ is actually two times smaller in magnitude than is commonly thought, because one should rely on the formula $F_T = 2\Phi_1,$ and not $F_T = \Phi_1.$

\section{Some applications of the hyperreactive model}
This section is not intended to give an extended and detailed list of all problems of mechanics of the variable-mass point solved using the proposed hyperreactive model. Our aim is to evaluate the new model and to substantiate the basic dynamic relations. Thus, as working formulas, we propose the solution algorithm of {\it Tsiolkovsky's problem in general formulation} for the hyperreactive case. 

Let a point move along a straight line in a non-resisting medium with no external forces applied. Evidently, the relative exhaust velocity of particles $V$ in this case is collinear to the reactive vector $R$ and directed opposite to the point motion. The law of motion and the velocity of the point are to be determined. Under the assumptions made, we obtain the integral of motion $Q(t) = C = {\rm const},$ which, in scalar form, yields the following equation:
$$M\,{{dR}\over{dt}} + {{dM}\over{dt}}\,R = C_0.\eqno(10)$$
Here
$$C_0 = M_0\,\biggl({{dR}\over{dt}}\biggr)_0 + \biggl({{dM}\over{dt}}\biggr)_0\,R_0,
\quad M_0 = M(t_0),$$
$$\biggl({{dR}\over{dt}}\biggr)_0 = \biggl({{dR}\over{dt}}\biggr)_{t=t_0},
\quad \biggl({{dM}\over{dt}}\biggr)_0 = \biggl({{dM}\over{dt}}\biggr)_{t=t_0},
\quad R_0 = R(t_0).$$

Integration of (10) with respect to $R$ yields
$$R = {{M_0R_0 + C_0\,(t - t_0)}\over M},\qquad {{dR}\over{dt}} = v - u,$$
where
$${{dR}\over{dt}} = {{C_0}\over M} - {{M_0R_0 + C_0\,(t - t_0)}\over M} \cdot {{
d\,{\rm ln}\,M}\over{dt}}.$$

It is of great interest to compare the obtained results with Tsiolkovsky's formula, which is substantiated based on additional assumptions about sta\-tion\-ari\-ty of the particle flow and invariability of the value $V$ (Tsiolkovsky's hypothesis). For this model, we have Meshchersky's equation
$$M\,{{dv}\over{dt}} = -\,V\,{{dM}\over{dt}},\qquad V = {\rm const},$$
from which we obtain Tsiolkovsky's logarithmic law (formula) after integration:
$$v = v_0 + V\,{\rm ln}\,{{M_0}\over M},\qquad v_0 = v(t_0).$$

Let us pay attention to incorrectness of these relations: on the one hand, $V = {\rm const},\ u = {\rm const}$, and hence $v = {\rm const}$; but, on the other hand, the speed $v$ changes according to the nonlinear logarithmic law. The stationarity is probably not assumed, though the central requirement $V = {\rm const}$ remains. But then there is a contradiction again, because the scalar (single-axis) equation for the old model should have the form
$$M\,{{dv}\over{dt}} = -\,V\,{{dM}\over{dt}} + M\,{{du}\over{dt}},\qquad V\,{{
dM}\over{dt}} = 0.$$

What conclusion can thus be drawn? In the context of the old model, Tsiolkovsky's hypothesis is not correct, and the requirement $V = {\rm const}$ cannot be ensured by the motion of the point. If we drop the excessive condition $V = {\rm const},$ then we get the integral of motion 
$${V\over{V_0}} = {{M_0}\over M},\qquad v = u + {{M_0V_0}\over M},$$
which is natural for the law $dQ_*(t)/dt = 0,\ Q_* = Q_1 - Q_2.$

Now let us consider the general non-stationary hyperreactive case. Though it might seem surprising, Tsiolkovsky's hypothesis is quite correct for the new model; moreover, it seems that Tsiolkovsky understood the reasonableness of his hypothesis and tried to make it fit to the not entirely appropriate model of Meshchersky. Let us suppose that vectors $V$ and $R$ are collinear, and $V = {\rm const} \not= 0:$
$$M\,{{dV}\over{dt}} = -\,2V\,{{dM}\over{dt}} + \bigl[\,R_0 - V\,(t - t_0)\,\bigr]
\,{{d^2M}\over{dt^2}},\quad {{dV}\over{dt}} = 0,\eqno(11)$$
where $dR/dt = - V$ is a given constant value.

From equation (11), we obtain the differential law of change of the mass that ensures fulfillment of the main requirement $V = {\rm const}:$
$${{d^2M}\over{dt^2}} - {{2V}\over{R_0 - V\,(t - t_0)}} \cdot {{dM}\over{dt}} =
0.\eqno(12)$$
Equation (12) is a total differential equation, i.e.
$$\bigl[\,R_0 - V\,(t - t_0)\,\bigr]\,{{dM}\over{dt}} - VM = C,\eqno(13)$$
where $C = R_0\,(dM/dt)_0 - VM_0.$
\smallskip

Integrating of relation (13) yields the dependence of mass as a function of initial data, time $t$, and the relative velocity $V:$
$$M(t) = {{R_0M_0 + C\,(t - t_0)}\over{R_0 - V\,(t - t_0)}}.$$

Now let us talk about the {\it Tsiolkovsky's second problem} in case his hypothesis is fulfilled: the problem of upward motion of a variable-mass point in the homogeneous gravity field. The equation for the specified constant relative velocity $V$ looks as follows:
$$M\,{{dV}\over{dt}} = -\,2V\,{{dM}\over{dt}} + \bigl[\,R_0 - V\,(t - t_0)\,\bigr]
\,{{d^2M}\over{dt^2}} - Mg,$$
where $dV/dt = 0,\ g$ is the gravity acceleration, from where we obtain the equation, according to which the change of mass should occur in order to ensure the fulfillment of the condition $V = {\rm const}$:
$${{d^2M}\over{dt^2}} - {{2V}\over{R_0 - V\,(t - t_0)}} \cdot {{dM}\over{dt}}
- {{gM}\over{R_0 - V\,(t - t_0)}} = 0.$$
Using standard methods, this equation is reduced to Riccati equation in one of non-integrable in closed form cases.

Consider the {\it exterior ballistics problem} using the new model under the following simplifying assumptions. We suppose that the reactive and hyperreactive forces are directed along the tangent to the trajectory, the gravity field is homogeneous, and "the Earth is flat". We also assume that the medium resistance force $G$ is directed along the tangent to the trajectory of motion of the object, and
$$G = {{L\rho S v^2}\over 2} = {{L\rho_0\rho Sv^2}\over{2\rho_0}},$$
where $L$ is the drag coefficient; $\rho,\,\rho_0$ is the air density at the altitude $z$ and near the Earth surface, respectively, $S$ is the characteristic area of the object, $v$ is the object velocity.

Let us assume that the coefficient $L$ can be represented in the following form: $L = P(v)\,H(z),$
where $P,H$ are continuous functions of respective arguments. Then
$$G = {{P(v)\,H(z)\,\rho_0 \rho S v^2}\over{2\rho_0}} = k\,h(z)\,\delta(v),$$
where the following notations are introduced:
$$h(z) = {{H(z)\,\rho}\over{\rho_0}},\qquad k = {{\rho_0 S}\over 2},\qquad \delta(v)
= P(v)\,v^2.$$

Based on the assumptions made about the acting forces, one can state that the trajectory of the system is a flat curve. Equations of motion in projections onto axes $Ox$ and $Oz$ are as follows:
$$M\,{{d^2x}\over{dt^2}} = -\,\biggl(k\,h(z)\,\delta(v) + 2V\,{{dM}\over
           {dt}} + R\,{{d^2M}\over{dt^2}} + M\,{{du}\over{dt}}\biggr)
           \,\cos \theta,$$
           $$M\,{{d^2z}\over{dt^2}} = -\,Mg \eqno(14)$$
           $$-\,\biggl(k\,h(z)\,\delta(v) + 2V\,{{dM}\over{dt}} + R\,{{d^2M}\over{dt^2}}
           + M\,{{du}\over{dt}}\biggr)\,\sin \theta.$$

Taking into account that $v \cos \theta = dx/dt,\ v \sin \theta = dz/dt,$ the system (14) can be represented in the following form:
$$vM\,{{d^2x}\over{dt^2}} = -\,\biggl(k\,h(z)\,\delta(v) + 2V\,{{dM}\over
           {dt}} + R\,{{d^2M}\over{dt^2}} + M\,{{du}\over{dt}}\biggr)\,{{dx}\over
           {dt}},$$
           $$vM\,{{d^2z}\over{dt^2}} = -\,Mg$$
           $$-\,\biggl(k\,h(z)\,\delta(v) + 2V\,{{dM}\over{dt}} + R\,{{d^2M}\over
           {dt^2}} + M\,{{du}\over{dt}}\biggr)\,{{dz}\over{dt}}.$$

Projecting the main vector equation (9) under the assumptions made onto the tangent to the trajectory and onto the normal to the trajectory, and complementing it with two kinematic relations, we obtain the following system of differential equations of motion:
$${{dv}\over{dt}} + g \sin \theta + {1\over M}\,\biggl(k\,h(z)\,\delta(v) + 2V
\,{{dM}\over{dt}} + R\,{{d^2M}\over{dt^2}} + M\,{{du}\over{dt}}\biggr) = 0,$$
$$v\,{{d\theta}\over{dt}} + g \cos \theta = 0,\quad {{dx}\over{dt}} - v \cos \theta
= 0,\quad {{dz}\over{dt}} - v \sin \theta = 0.\eqno(15)$$

We hardly need to comment the system (15), which is well known without hyperreactive components [19]. As before, the principal goal in using the new model for the ballistic problem is to write down accurate dynamic relations taking into account hyperreactive terms.

Let us also note that the {\it equations of ballistic motion} (15) can be solved using a computer, and the numerical analysis will help thoroughly determine all specific features of motion. The proposed hyperreactive model, where the mass is one more independent variable, allows for reducing errors and deviations of ballistic trajectories to a minimum, because in this case the object dynamic is described by the most accurate equations of motion.

One of the most important conclusions from the presented theory: the momentum of variable-mass dynamic systems includes all velocity characteristics, namely changes of position and changes of mass, since in such systems the mass gets all attributes of a new independent variable.

\section{Tsiolkovsky's problems in hyperreactive\\ formulation}

The most important conclusion from the presented hyperreactive theory is that this method shows the main directions for obtaining additional forces whereby the original dynamic system can reach very high absolute velocities. 

The validity of the new approach can be verified experimentally and confirmed in practice. The advantages of the new method can be set forth in the following statements: the calculation of curves allows for the minimization of errors and correction of trajectories, as well as for the optimal selection of the fuel reserve (of a rocket) during the entire motion process.

Before, based on the statement of independence of generalized co\-or\-dinates of a point and its mass $M(t)$ at a current moment of time $t$, $t \in [\,t_0,\,t\,],$ and assuming that the mass is one more generalized (Lagrange) coordinate, the total momentum of the point $Q(t)$ was calculated according to the formula
$$Q(t) = M(t)\,{{dR(t)}\over{dt}} + {{dM(t)}\over{dt}}\,R(t),$$
where the reactive vector $R(t) = r(t) - \rho(t),$\ \ and  $dr(t)/dt = v(t),$ $d\rho(t) /dt = u(t).$

The (symmetrical) momentum introduced in this way allows obtaining the new differential law of dynamics (principle of completeness):
$${{dS(t)}\over{dt}} = Q(t)$$
for the motion composition vector $S(t) = M(t)\,R(t).$

Then, applying the law of changing the momentum to the value $Q(t),$ we can write down the equation of hyperreactive motion of the point (9):
$$M\,{{dv}\over{dt}} = F + M\,{{du}\over{dt}} - 2\,{{dM}\over{dt}}\,{{dR}\over
{dt}} - {{d^2M}\over{dt^2}}\,R.$$

Based on the new concept of motion of a variable-mass point, several problems associated with the particular conditions of realization of such motion are considered below, and characteristic features that relate to the principle of completeness are identified.
\medskip

{\bf 3.1. The Tsiolkovsky's first problem.} Let us study in more detail the {\it Tsiolkovsky's first problem}, when the variable-mass point moves along a straight line in a non-resisting medium with no external forces applied under the assumption that the relative exhaust velocity of particles $V$ is constant, collinear to the reactive vector of motion $R$, and directed opposite to the point motion. Under these conditions, which constitute the Tsiolkovsky's hypothesis, it is required to determine the motion of the point.

It was already mentioned above that Tsiol\-kov\-sky's logarithmic law is contradictory.  Therefore, we use the scalar equation of hyperreactive motion (9). Under the assumptions made, we obtain
$${{dR}\over{dt}} = -\,{{R\,d^2M/dt^2}\over{2\,dM/dt}}.\eqno(16)$$

The integration of (16) with respect to $R$ yields
$$R = R_0\,\exp\,\biggl\{\,-\,{1\over 2}\,{\rm ln}\,\biggl[\,{{dM/dt}\over{(dM/dt)_0}}
\,\biggr]\,\biggr\},$$
where $R_0 = R(t_0),\ (dM/dt)_0 = (dM/dt)_{t=t_0},$ from where
$$\biggl({R\over{R_0}}\biggr)^2 = \biggl({{dM}\over{dt}}\biggr)_0\ \bigg/\ {{dM}\over
{dt}}.\eqno(17)$$
Substitute the obtained expression (17) into relation (16). Then
$${{dR}\over{dt}} = -\,{{R_0\,(dM/dt)_0^{1/2}\ (d^2M/dt^2)}\over{2\,(dM/dt)^{3/2}}}.
\eqno(18)$$

Let us assume for definiteness that the point mass $M(t)$ is a continuous, monotonically decreasing function of time:
$${{dM}\over{dt}} < 0,\qquad {{d^2M}\over{dt^2}} > 0,\qquad {{(dM/dt)_0}\over{
dM/dt}} > 0.$$
Hence, the right-hand side of (18) is a positive value, and the point velocity $v$ increases according to the following law:
$$v = u - {{K_0\,d^2M/dt^2}\over{(dM/dt)^{3/2}}},\eqno(19)$$
where $K_0 = R_0\,(dM/dt)_0^{1/2}\,/\,2.$

Compliance with dependence (19) is ensured by the change of mass according to a certain given law $M = M(t).$ The point velocity $v(t)$ will be determined if the velocity $u(t)$ in relation (19) is a known function of time. If relative velocity $V = {\rm const}$ is also given, the change of mass should be governed by the following differential rule:
$$K_0\,{{d^2M}\over{dt^2}} = V\,\biggl({{dM}\over{dt}}\biggr)^{3/2}.$$

This equation can be represented in the following form:
$${{d^2M}\over{dt^2}} - {{2V}\over{R_0 - V\,(t - t_0)}} \cdot {{dM}\over{dt}} = 0,$$
from where we obtain, after integrating, the law of change of the mass:
$$M(t) = {{R_0M_0 + C\,(t - t_0)}\over{R_0 - V\,(t - t_0)}},\quad C \equiv R_0
\,\biggl({{dM}\over{dt}}\biggr)_0 - VM_0,$$
or
$$M(t) = M_0 + {{R_0\,(dM/dt)_0\,(t - t_0)}\over{R_0 - V\,(t - t_0)}},\quad
\biggl({{dM}\over{dt}}\biggr)_0 < 0.$$

Now let us consider some {\it laws of mass change}. The law of mass change is determined by the mode of operation of the radiating center (engine). Let us find out for which changes of mass $M(t)$ the constancy of reactive and hyperreactive forces is ensured in the case of rectilinear motion. Let
$$-\,{{dM(t)}\over{dt}} {{dR(t)}\over{dt}} = C_r,\eqno(20)$$
where $C_r = {\rm const} \not= 0.$ As $dR(t)/dt = -\,V(t),$ integrating relation (20) over the time interval $[\,t_0,\,t\,],$ we get
$$M(t) = M_0 + C_r \int_{t_0}^t {{ds}\over{V(s)}}.\eqno(21)$$

Let us discuss formula (21). In this relation, $V(t) \not= {\rm const}.$ Indeed, if we assume the contrary, we obtain the linear law of mass change, from where $d^2M(t)/dt^2 = 0.$ In the main equation of dynamics (9) $dV/dt = 0$ in the left-hand side; if condition $F = 0$ is fulfilled, we reach a contradiction, and Tsiolkovsky's hypothesis is inconsistent.

Consider the case when
$$-\,{{d^2M(t)}\over{dt^2}}\,R(t) = C_h,\eqno(22)$$
where $C_h = {\rm const} \not= 0,\ R(t) \not= {\rm const}.$ Having integrated equation (22) twice, we obtain the dependency
$$M(t) = M_0 + \biggl({{dM}\over{dt}}\biggr)_0 (t - t_0) - C_h \int_{t_0}^t \int_{t_0}^s
{{dw\,ds}\over{R(w)}}.$$

Since many papers take a special interest in two cases of mass change,
1) linear law $M(t) = M_0\,[\,1 -$ $- \alpha\,(t - t_0)\,]$ and 2) exponential law $M(t) = M_0\,\exp\,[\,- \alpha\,(t - t_0)\,],$ where $\alpha > 0,$ we give particular attention to them.

If $M(t) = M_0\,[\,1 - \alpha\,(t - t_0)\,],$ then the mass flow per second equals to 
$dM/dt = -\,\alpha M_0,$ where the parameter $\alpha$ is called the specific mass flow per second; besides, $d^2M/dt^2 = 0.$ Let $F = 0.$ Then Tsiolkovsky's hypothesis about the constant $V$ is not fulfilled, and the relative velocity for the linear case should satisfy the equation
$$M\,{{dV}\over{dt}} = -\,2V\,{{dM}\over{dt}},$$
i.~e.
$$V(t) = {{V_0}\over{\bigl[\,1 - \alpha\,(t - t_0)\,\bigr]^2}},\qquad R(t) = R_0
- {{V_0\,(t - t_0)}\over{1 - \alpha\,(t - t_0)}},\eqno(23)$$
where $V_0 = V(t_0).$ Let $a_r$ denote the acceleration induced by the action of the double reactive force. Then, in the case of the linear law of mass change, we have with the consideration of (23):
$$a_r = {{2V\,dM/dt}\over M} = -\,{{2\alpha V_0}\over{\bigl[\,1 - \alpha\,(t - t_0
)\,\bigr]^3}}.$$
The overload produced by the reactive force is, in this case, equal to
$$n = {{a_r}\over g} = -\,{{2\alpha V_0}\over{g\,\bigl[\,1 - \alpha\,(t - t_0)
\,\bigr]^3}},$$
where $g$ is the gravity acceleration.

Now consider the exponential law of mass change $M(t) = M_0\,\exp\,[\,
- \alpha\,(t - t_0)\,].$ We have
$${{dM}\over{dt}} = -\,\alpha M_0\,e^{- \alpha\,(t - t_0)} = -\,\alpha M,\quad
{{d^2M}\over{dt^2}} = \alpha^2 M.$$
Let $F = 0.$ Notice that Tsiolkovsky's hypothesis is not fulfilled for the exponential law due to the inconsistency of equations either. We have 
$$M\,{{d^2R}\over{dt^2}} = -\,2\,{{dM}\over{dt}}\,{{dR}\over{dt}} - {{d^2M}\over{
dt^2}}\,R,$$
or
$${{d^2R}\over{dt^2}} - 2\alpha\,{{dR}\over{dt}} + \alpha^2 R = 0.\eqno(24)$$

After integrating the equation (24), we obtain:
$$V(t) = \bigl[\,V_0 + \alpha\,(V_0 + \alpha R_0)(t - t_0)\,\bigr]\,\exp\,[\,
\alpha\,(t - t_0)\,],$$
$$R(t) = \bigl[\,R_0\ -\ (V_0 + \alpha R_0)(t - t_0)\,\bigr]\,\exp\,[\,\alpha\,(
t - t_0)\,],$$
from where we get expressions for accelerations and overload:
$$a_r = -\,2\alpha V,\quad a_h = -\,\alpha^2 R,\quad a_{r+h} = -\,\alpha\,(2V +
\alpha R),\quad n = {{a_{r+h}}\over g},$$
where $a_{r+h} = a_r + a_h.$
\medskip

{\bf 3.2. The Tsiolkovsky's second problem.} Now let us study {\it the second Tsiolkovsky problem}, namely the vertical upward motion of a variable-mass point in a homogeneous gravity field. It is necessary to determine the law of the change of speed and distance as functions of time and to find the maximum altitude of the point. The relative velocity $V$ of emitted particles is constant and directed vertically downwards. The equation of motion in this case takes the following form:
$${{dR}\over{dt}} = -\,{{R\,d^2M/dt^2}\over{2\,dM/dt}} - {{Mg}\over{2\,dM/dt}}.
\eqno(25)$$

By integrating equation (25) we obtain:
$$R(t) = {{(dM/dt)_0^{1/2}}\over{(dM/dt)^{1/2}}}\ \biggl[\,R_0 - {g\over{2\,(
dM/dt)_0^{1/2}}}\,\int_{t_0}^t {{M\,dt}\over{(dM/dt)^{1/2}}}\,\biggr].\eqno(26)$$
Substituting (26) into expression (25), we find the dependence $dR(t)/dt$ in function $M(t),\,dM(t)/dt$ and $d^2M(t)/dt^2,$ from where
$$v = u - {{gM}\over{2\,dM/dt}} \eqno(27)$$
$$-\ {{(dM/dt)_0^{1/2}\,d^2M/dt^2}\over{2\,(dM/dt)^{3/2}}}\ \biggl[\,R_0 -
{g\over{2\,(dM/dt)_0^{1/2}}}\,\int_{t_0}^t {{M\,dt}\over{(dM/dt)^{1/2}}}\,\biggr],$$
where $u(t),\,M(t)$ are given functions of time.

If the constant relative velocity $V$ is specified, then equation (25) represents
the differential law of mass change in the form of
$${{d^2M}\over{dt^2}} - {{2V}\over{R_0 - V\,(t - t_0)}} \cdot {{dM}\over{dt}} -
{{gM}\over{R_0 - V\,(t - t_0)}} = 0.\eqno(28)$$

As is well known, the second order linear homogeneous equation (28) is not integrable in the closed form. Note that equation (28) represents the equation of relative equilibrium of a variable-mass point in gravity field under the assumption that $V = u = {\rm const}.$

Now let us find the maximum altitude $H$ of the point. As $v(t_*) = 0$ for this altitude, the time of rising $t_*$ can be determined from equation (27). Then we have (for $H(t_0) = 0):$
$$H = \int_{t_0}^{t_*} v(s)\,ds = R(t_*) - R(t_0) + \int_{t_0}^{t_*} u(s)\,ds$$
$$=\ R(t_*) + \rho_0 + \int_{t_0}^{t_*} u(s)\,ds,\qquad \rho_0 = \rho(t_0),$$
where $u(t),\,M(t)$ are given time functions, and the value $R(t_*)$ is determined using the formula (26) for $t = t_*.$

Comparing the obtained result with the case when $M(t) = M_0 \, \exp\,[\,- \alpha\,
(t - t_0)\,],\,\alpha > 0,$ $V \not= {\rm const},$ we get
$$M\,{{d^2R}\over{dt^2}} = -\,2\,{{dM}\over{dt}}\,{{dR}\over{dt}} - {{d^2M}\over{
dt^2}}\,R - Mg$$
or
$${{d^2R}\over{dt^2}} - 2\alpha\,{{dR}\over{dt}} + \alpha^2R = -\,g,$$
from where we find
$$R(t) = \bigl[\,R_0 - (V_0 + \alpha R_0)(t - t_0)\,\bigr]\,\exp\,[\,
         \alpha\,(t - t_0)\,] - {g\over{\alpha^2}},$$
         $$V(t) = \bigl[\,V_0 + \alpha\,(V_0 + \alpha R_0)(t - t_0)\,\bigr]
         \,\exp\,[\,\alpha\,(t - t_0)\,].\eqno(29)$$
Time $t_*$ may be obtained from the equation
$$u(t_*) = \bigl[\,V_0 + \alpha\,(V_0 + \alpha R_0)(t - t_0)\,\bigr]\,\exp\,[\,
           \alpha\,(t_* - t_0)\,],$$
and then the value of $H$ is determined. The case when the mass decreases according to the linear law can be considered in a similar way.
\medskip

{\bf 3.3. Optimal conditions of motion.} Now let us proceed with the investigation of optimal conditions of motion in Tsiolkovsky problems. We have shown that the main integral characteristics of the motion of a point depend on the law of the change of its mass; therefore, we have a reason to speak about the formation of optimal conditions of motion.

Let $m(t)$ denote the fuel reserve: $M(t) = N + m(t),$ where $N = {\rm const}$ is the point mass without fuel. From relations (19) and (27), it follows that the velocity acquired by the point depends on the speed $(M(t)/dt = dm(t)/dt)$ and acceleration  $(d^2M(t)/dt^2 = d^2m(t)/dt^2)$ of the change of the fuel mass. This is a very important conclusion.

Expression (19) does not contain mass at all. Let us try to forecast this result from logical considerations. According to the earlier (consumption based) theory, the greater is the mass of fuel of the object, the greater velocity it acquires; besides, the nature of the change of values $dm(t)/dt$ and $d^2m(t)/dt^2.$ does not matter.

In reality, it is far from truth, of course: taking into account the mass of fuel while ignoring these values actually means taking into account only impact effects during the motion of the object (meaning the non-concentrated or {\it distributed impact}). In reality, as the new theory shows, even a small mass of fuel in the case of corresponding changes can provide for considerable and required absolute velocities.

Let us return to the Tsiolkovsky's first problem. The length of the active section, when the mass of fuel $m(t)$ becomes zero at $t = t_*,$ will be equal to 
$$r(t_*) = \rho_0 + \int_{t_0}^{t_*} u(s)\,ds + {{R_0\,(dM/dt)_0^{1/2}}\over{(
dM/dt)_*^{1/2}}},$$
where time $t_*$ is determined from the equation $M(t_*) = N.$ If the relative velocity $V = {\rm const}$ is specified, then $t_*,\ (dM/dt)_* = (dM/dt)_{t=t_*}$ are calculated using the earlier derived formula
$$M(t) = {{R_0M_0 + C\,(t - t_0)}\over{R_0 - V\,(t - t_0)}}.$$

Thus, we have
$$t_* = t_0 + {{m_0R_0}\over{m_0V - R_0\,(dm/dt)_0}},$$
$$\biggl({{dM}\over{dt}}\biggr)_* = \biggl({{dm}\over{dt}}\biggr)_* = {{R_0^2\,
(dm/dt)_0}\over{\bigl[\,R_0 - V\,(t_* - t_0)\,\bigr]^2}}.$$

The case of an instantaneous fuel ejection (a concentrated single impact) corresponds to $t_* = t_0,$ i.$\,$e.
$$r(t_*) = r(t_0) = \rho_0 + R_0,\quad {V\over{R_0}} - {{(dm/dt)_0}\over{m_0}}
\to \infty.$$
If
$$V \to {{R_0\,(dm/dt)_0}\over{m_0}},$$
then $t_* \to \infty:$ an infinitesimal change of the fuel mass is observed and, in addition, $r(t_*) \to \infty.$

Now let us proceed to the Tsiolkovsky's second problem. Let us find the maximum altitude of the active section for the rising of a variable-mass point that moves vertically upwards in a homogeneous gravity field in the case when the law of changing $M(t)$ is given in advance. We have $M(t_*) = N$ (for $V = {\rm const},$ and $V$ is not given a priori):
\smallskip
$$R(t_*) = {{(dm/dt)_0^{1/2}}\over{(dm/dt)_*^{1/2}}}\ \biggl[\,R_0 - {g\over{2\,
(dm/dt)_0^{1/2}}}\,\int_{t_0}^{t_*} {{M\,dt}\over{(dm/dt)^{1/2}}}\,\biggr].$$
\smallskip

This problem will have an analytical solution if the function of time $M(t)$ is known.
Let
$$M(t) = M_0\,\exp\,[\,-\,\alpha\,(t - t_0)\,],\quad M_0 = N + m_0,\quad V \not=
{\rm const}.$$
Solving the equation $M(t_*) = N,$ we can find the moment of time $t_*,$ at which fuel reserve will be equal to zero (end of the active section):
$$(N + m_0)\,\exp\,[\,-\,\alpha\,(t_* - t_0)\,] = N,$$
from where
$$t_* = t_0 + {\rm ln}\ \biggl(1 + {{m_0}\over N}\biggr)^{1/\alpha}.\eqno(30)$$

Consider equation (29) for $R(t).$ Substitution of expression (30) into relation (29) yields
$$R(t_*) = \biggl[\,R_0 - (V_0 + \alpha R_0)\,{\rm ln}\,\biggl(1 + {{m_0}\over
N}\biggr)^{1/\alpha}\,\biggr]\biggl(1 + {{m_0}\over N}\biggr) - {g\over{\alpha^2}}.
\eqno(31)$$
Substituting relations (30) and (31) into the expression for
$$H = \rho_0 + \int_{t_0}^{t_*} u(s)\,ds + R(t_*),$$
one gets the value of the altitude of the active section for the case when fuel reserve decreases according to the exponential law.

Determine the value of $\alpha$ for which the altitude of the rising of the point in this case is maximum. Differentiating $H$ with respect to $\alpha,$ we bring the equation $\partial H/\partial \alpha = 0$ for the determination of the optimal value of $\alpha$ (after reduction by $1/\alpha^2)$ to the form
$${\rm ln}\,\biggl(1 + {{m_0}\over N}\biggr)\,u \biggl[\,t_0 + {\rm ln}\,
\biggl(1 + {{m_0}\over N}\biggr)^{1/\alpha}\,\biggr] = {{2g}\over \alpha} +
V_0\,\biggl(1 + {{m_0}\over N}\biggr)\,{\rm ln}\,\biggl(1 + {{m_0}\over N}\biggr).
\eqno(32)$$

In particular, the maximum altitude of the active section can be reached if one assumes $\alpha = \infty$ in equation (32), which corresponds to the instantaneous ejection of fuel. This imposes a restriction for the choice of initial data: the condition $u(t_0) = V_0\,(1 + m_0/N)$ should be satisfied.

We would like to emphasize the fact that the earlier theory determines the maximum altitude only for $\alpha = \infty;$ the new hyperreactive model provides for a whole family of  solutions of this problem. The case of a linear decrease of the fuel mass can be considered in the same way, and we will not delve into it.

If the value $V = {\rm const}$ is specified in advance, then equation (28) is not integrable in closed form with respect to $M(t).$ However, by solving it numerically, it is possible to approximately determine the time $t_*,$ for which $m(t_*) = 0;$ thus, the maximum altitude of the active section will be found.
\medskip

{\bf 3.4. Some conclusions.} Let us formulate the main conclusions based on the above results obtained with hyperreactive modeling, though it is clear that their number could be much larger.
\smallskip

1. {\it The Meshchersky-Tsiolkovsky model} is flawed in its conceptual foundations. Firstly, it is contradictory at the differential level; secondly, it does not allow for the consideration of terms containing $d^2M(t)/dt^2$; finally, this model does not take into account the influence of the mass of the system on the magnitude of its momentum.

2. {\it The hyperreactive model} allows for achieving the required absolute velocities mainly due to the nature of the velocity and acceleration of the object mass change.

3. The previous theory puts emphasis on setting the relative velocity of particle outflow $V,$ while the new one gives preference to setting the absolute velocity of particle outflow $u.$ In this case, the dynamics of the object is completely defined.

4. As the obtained results show, Tsiolkovsky's assumption that $V = {\rm const}$ is not justified in many cases: the point motion in this situation has "constrained" dynamics.

5. For efficient functioning of the hyperreactive model and for attaining the required final velocity, it is necessary to ensure the acceleration of fuel particles (or, better to say, not fuel but working substance) inside the engine, a kind of a compact accelerator [9, 10].
\medskip

The hyperreactive engine should provide for the accelerated motion of particles of the working substance. The design of such an accelerating engine--accelerator might be based on the most diverse physical and technical ideas, for example, related with the implementation of the project of the nuclear electric generator $\hbox{(see part III}$ of book [10]). At the current stage of the development of the rocket technology, a small hyperreactive effect in real systems manifests itself in the occurrence of non-stationarity (during the boost phase) and in a deviation of the trajectory from calculated one.

\section{Energy of a variable-mass point. Hamilton's\\ variational principle}
In the previous part of the article, the general problem of the derivation of equations of dynamics of a point subjected to the change of mass as a function of mass itself, as well as of the velocity and acceleration of its change depending on time, was formulated and solved. In spite of the obvious importance of such a dynamic study, the issues of power support of the hyperreactive motion and its fundamental relation with variational principles of mechanics remained out of the scope of the analysis. The second part of the article deals with resolving these issues.

For the sake of simplicity, we will provide reasoning for an individual point the mass of which changes with time, and there are no constraints imposed on its position. Summing up all expressions for points, one can finally obtain the hyperreactive transformations for a mechanical system of material points and further for a variable-mass body.

Thus, a point with mass $M(t)$ and external force $F(t)$ acting on it is described by the equation of motion (9):
$${{d\,Q(t)}\over{dt}} = F(t),\qquad Q(t) = M(t)\,\dot{R}(t) + \dot{M}(t)\,R(t),$$
where $Q(t)$ is the total (symmetrical) momentum of the point, and $R(t)$ is the reactive vector of motion.

Let the motion of the point be not constrained by geometrical (holonomic or non-holonomic) constraints. Choose the coordinates of vector $R(t)$ as generalized coordinates characterizing the position of the point in space. We will attempt to find out the energy of motion of the point in the hyperreactive case, that is, for dynamical description using equation (9).

We have the following chain of identical transformations (under the assumption that the point mass $M(t)$ is a continuously differentiable function of only time $t):$
$$F = {d\over{dt}}\,\bigl(M\,\dot{R} + \dot{M}\,R\bigr) = {d\over{dt}}\,\bigl(M\,
\dot{R} + \dot{M}\,R\bigr)\cdot {{\partial R}\over{\partial R}}$$
$$=\ {d\over{dt}}\,\biggl[\,\bigl(M\dot{R} + \dot{M} R\bigr)\,{{\partial R}\over
{\partial R}}\,\biggr] - \bigl(M\dot{R} + \dot{M}R\bigr)\,{d\over{dt}}\,\biggl(
{{\partial R}\over{\partial R}}\biggr) \eqno(33)$$
$$=\ {d\over{dt}}\,\biggl[\,\bigl(M\dot{R} + \dot{M} R\bigr)\,{{\partial\dot{R}}
\over{\partial\dot{R}}}\,\biggr] = {d\over{dt}}\,{\partial\over{\partial\dot{R}}}
\,\biggl({{M\dot{R}^2}\over 2} + \dot{M} R\dot{R}\biggr),$$
where
$${{\partial R}\over{\partial R}} = {{\partial\dot{R}}\over{\partial\dot{R}}} =
1,\qquad {d\over{dt}}\,\biggl({{\partial R}\over{\partial R}}\biggr) = 0.$$

To simplify writing, hereinafter we will omit the word "symmetrical" in the phrase "total symmetrical momentum" where applicable. Thus, using relation (33), {\it the efficient energy} $T_e$ of point motion in the field of the external force $F$ can be represented in the form of the following sum:
$$T_e = T_k + T_r,\eqno(34)$$
where the following values are denoted in terms of the reactive vector $R$: $T_k = M\dot{R}^2/2$ is {\it kinetic energy}, $T_r = \dot{M} R\dot{R}$ is the newly introduced {\it reactive energy} of the point.

Along with the theorem of changing of the total momentum of a point, based on relations (33) and (34), we will give the theorem of changing of the efficient energy of a variable-mass point.

\begin{theorem}
Time derivative of the gradient vector of the efficient energy scalar field $T_e$ $(34)$ of a point with respect to the elements of vector $\dot{R}$ $($or negative relative velocity $-\,V$ of particle outflow$)$ is equal to the vector of the external active force acting on the point:
$${d\over{dt}}\,\bigl(\nabla_{\dot{R}}\ T_e\bigr) = F.\eqno(35)$$
\end{theorem}

{\bf Remark.} In equation (35), $\nabla_{\dot{R}} = \partial/\partial \dot{R}$ denotes the vector of partial differentiation (gradient) with respect to the corresponding components of the vector $\dot{R}.$ This allows us to conclude that the point has the highest rate of change of the efficient energy in the direction of the velocity vector $\dot{R}.$
\smallskip

Let the position of the system of $n$ points with masses $M_i(t)$ and forces $F(t) = \bigl(F_i(t) \bigr),\,i = \overline{1,n}$ be defined by $s$ independent generalized coordinates
$q(t) = \bigl(q_i(t)\bigr),\,j = \overline{1,s},$ and holonomy takes place:
$$R(t) = \bigl(R_i(t)\bigr) = R(q(t)),\eqno(36)$$
where vectors $F_i(t),\,R_i(t)$ represent the force vector and the reactive vector in three-dimensional Euclidean space $R^3$ for the $i^{th}$ point, respectively.

Let us use the known kinematic relations for curvilinear and Cartesian coordinates and their velocities:
$${{\partial R_i}\over{\partial q_j}} = {{\partial\dot{R}_i}\over{\partial\dot{q}_j}},
\qquad {d\over{dt}}\,\biggl({{\partial R_i}\over{\partial q_j}}\biggr) =
{{\partial\dot{R}_i}\over{\partial q_j}}.\eqno(37)$$

One can assign to the coordinate $q_j$ the generalized external force
$$F_i\,{{\partial R_i}\over{\partial q_j}} = G_j,\eqno(38)$$
where summation over the same indices is performed in (37) and everywhere below.

Taking into account relations (36) and (37), we have:
$${d\over{dt}}\,\bigl(M_i\dot{R}_i + \dot{M}_iR_i\bigr)\cdot {{\partial R_i}\over
{\partial q_j}}$$
$$=\ {d\over{dt}}\,\biggl[\,\bigl(M_i\dot{R}_i + \dot{M}_iR_i\bigr)\,{{\partial
R_i}\over{\partial q_j}}\,\biggr] - \bigl(M_i\dot{R}_i + \dot{M}_iR_i\bigr)\,
{d\over{dt}}\,\biggl({{\partial R_i}\over{\partial q_j}}\biggr)$$
$$=\ {d\over{dt}}\,\biggl[\,\bigl(M_i\dot{R}_i + \dot{M}_iR_i\bigr)\,{{\partial
\dot{R}_i}\over{\partial\dot{q}_j}}\,\biggr] - \bigl(M_i\dot{R}_i + \dot{R}_i +
\dot{M}_iR_i\bigr)\,{{\partial\dot{R}_i}\over{\partial q_j}} \eqno(39)$$
$$=\ {d\over{dt}}\,{\partial\over{\partial\dot{q}_j}}\,\biggl({{M_i\dot{R}_i^2}
\over 2} + \dot{M}_iR_i\dot{R}_i\biggr) - {\partial\over{\partial q_j}}\,\biggl(
{{M_i\dot{R}_i^2}\over 2} + \dot{M}_iR_i\dot{R}_i\biggr)$$
$$+\ \dot{M}_i\dot{R}_i\,{{\partial R_i}\over{\partial q_j}} = {d\over{dt}}\,
{{\partial T_e}\over{\partial\dot{q}_j}} - {{\partial T_e}\over{\partial q_j}} +
\dot{M}_i\dot{R}_i\,{{\partial R_i}\over{\partial q_j}},$$
where $T_e$ is the efficient energy of the system of hyperreactive points.

Under the assumption that constraints (36) are ideal, one can derive from relation (39) the Lagrange equations of the second kind for the system of variable-mass points in the general hyperreactive case:
$${d\over{dt}}\,{{\partial T_e}\over{\partial\dot{q}_j}} - {{\partial T_e}\over
{\partial q_j}} = G_j + P_j,\qquad P_j = -\,\dot{M}_i\dot{R}_i\,{{\partial R_i}
\over{\partial q_j}},\eqno(40)$$
where $P_j$ is a generalized hyperreactive force corresponding to coordinate $q_j.$

Under assumption that the field of acting forces is potential, that is,
$$G = {\rm grad}\,U,\qquad G_j = {{\partial U}\over{\partial q_j}},$$
where $U(q_1,q_2,\,...,\,q_s)$ is {\it potential function}, $-\,U = \Pi$ is {\it potential energy}, we can write down the following for the {\it total kinetic potential} $T_*$ of the system:
$$T_* = T_e + U = L + T_r,\eqno(41)$$
where $L = T_k + U$ is the Lagrange function.

As the potential $U$ is a function of coordinates only, we have
$${{\partial T_*}\over{\partial\dot{q}_j}} = {{\partial T_e}\over{\partial\dot{q}_j}}.$$
Hence, in terms of the function $T_*$ (41), equation (40) can be written in the following form:
$${d\over{dt}}\,{{\partial T_*}\over{\partial\dot{q}_j}} - {{\partial T_*}\over
{\partial q_j}} = P_j.\eqno(42)$$

Now let us choose the functional $S_H$, which is called the Hamiltonian action, as the measure of mechanical motion. We derive Hamilton's variational principle from the equation of hyperreactive motion of a variable-mass point and establish the extremal properties of the action $S_H$ for real motions. For that, we will use known concepts and constructions of the calculus of variations for the case of synchronous variation of trajectories [20].

Thus, let a vector universal equation of hyperreactive motion be given, and
$$\biggl(F - {{d\,(M\dot{R} + \dot{M}R)}\over{dt}}\biggr)\,\delta R = 0,\eqno(43)$$
where $\delta R$ denotes the variation of the reactive vector. We suppose that the field of acting forces is potential, i.e.
$$F\,\delta R = {\rm grad}\,U\ \delta R = \delta U.\eqno(44)$$

For other terms in equation (43), the following transformations can be performed:
$${d\over{dt}}\,(M\dot{R} + \dot{M}R)\cdot \delta R$$
$$=\ {d\over{dt}}\,\bigl[\,(M\dot{R} + \dot{M}R)\,\delta R\,\bigr] - (M\dot{R} +
\dot{M}R)\,{d\over{dt}}\,(\delta R)$$
$$=\ {d\over{dt}}\,\bigl[\,(M\dot{R} + \dot{M}R)\,\delta R\,\bigr] - (M\dot{R} +
\dot{M}R)\,\delta\dot{R} \eqno(45)$$
$$=\ {d\over{dt}}\,\bigl[\,(M\dot{R} + \dot{M}R)\,\delta R\,\bigr] - \delta\,\biggl(
{{M\dot{R}^2}\over 2} + \dot{M}R\dot{R}\biggr) + \dot{M}\dot{R}\,\delta R,$$
where $M\dot{R}^2/2 + \dot{M}R\dot{R} = T_e.$ Here, $M(t)$ is a known continuously differentiable function of time and $M(t),\,\dot{M}(t)$ are not varied. Thus, expression (45) is written as
$${{dQ}\over{dt}}\cdot \delta R = {d\over{dt}}\,(Q\,\delta R) - \delta T_e +
\dot{M}\dot{R}\,\delta R.\eqno(46)$$

Taking into account relations (44) and (46), the universal equation of hyperreactive motion (43) can be represented in the following form:
$$\delta T_e + \delta U = {d\over{dt}}\,(Q\,\delta R) + \dot{M}\dot{R}\,\delta R$$
or
$$\delta T_*\,dt = d\,(Q\,\delta R) + \dot{M}\dot{R}\,\delta R\,dt.$$

Let us integrate the latter expression with respect to time from $t_0$ to $t_*.$ We obtain
$$\int_{t_0}^{t_*}\,\delta T_*\,dt = Q\,\delta R\,\big|_{t_0}^{t_*} +
\int_{t_0}^{t_*}\,\dot{M}\dot{R}\,\delta R\,dt.$$
In case of synchronous variation, for fixed time moments $t = t_0$ and $t =t_*$ we have $\delta R = 0,$ i.~e.,
$$\int_{t_0}^{t_*}\,\bigl(\delta T_* - \dot{M}\dot{R}\,\delta R\bigr)\,dt = 0.
\eqno(47)$$
Thus, Hamilton's principle for hyperreactive motion has a mathematical statement in the form of (47).

Hamilton's principle in the classic formulation is as follows: from all possible (with the consideration of imposed constraints) motions of a conservative mechanical system that transfer the system within a specified time from the given initial configuration into another given configuration, the actual motion under the action of given forces and reactions of constraints (ideal, holonomic, stationary) will be the motion for which the Hamiltonian action functional $S_H$ has an extreme value 
$$\delta\,\int_{t_0}^{t_*}\,L\,dt = \delta S_H = 0,\eqno(48)$$
where $L$ is the classical Lagrange function that represents the difference between the kinetic and potential energy of the system.

Hamilton's principle (48) can be formulated more briefly [21]: the Hamiltonian action $S_H$ has a stationary value if $\delta S_H = 0.$ It is also proven (see [14, 22]) that the action $S_H$ in this case takes not only a stationary value but also has only a minimum.

The compact formulation of Hamilton's principle in the form of relation (48) is very convenient, as it allows for relating the actual motion of the system with extremal properties of the Lagrange function. 
It is impossible to perform similar transformations in formula (47) using conventional methods of integral calculus and calculus of variations, as it is impossible to take the variation outside the integral in the second term of the expression under integral sign (47) in the same manner as it is done for the first term.

\section{Variational integral: structure and properties}
In order to write down formula (47) in the standard form, we proceed as follows. We introduce the concept of {\it variational integral} as a mathematical operation inverse to the operation of variation of a functional (the variational integral was defined for the first time in the paper [9]).

For this, we define a sufficiently smooth function $g(t)$ from $C^r[\,t_0,t_*\,],\,r \ge 1,$ on the interval $[\,t_0,t_*\,]$. Consider an admissible function (curve) $\bar{g}(t)$ which is close (in the sense of proximity of $r^{th}$ order) to the function $g(t)$.

As is well known, the difference $\delta\,g(t) = \bar{g}(t) - g(t)$ is called a variation of the function $g(t)$, and its role in the calculus of variations is similar to the role of the increment of the independent variable $\Delta t = dt$ in the problems of the differential calculus and study of the extrema of functions $g(t).$

In the interval $[\,t_0,t_*\,]$, we will also consider the functionals $f[\,g(t)\,]$ and $F[\,g(t)\,].$ Let us define the variation of the functional $\delta\,F[\,g(t)\,]$ in the conventional manner [20] as linear with respect to $\delta g$ part of increment of the functional $\Delta F,$ namely set
$$\delta\,F[\,g(t)\,] = L[\,g(t),\,\delta g\,],$$
where the functional increment
$$\Delta F = F[\,g(t) + \delta g\,] - F[\,g(t)\,]$$
can be represented in the following form:
$$\Delta F = L[\,g(t),\,\delta g\,] + \beta\,\big(g(t),\,\delta g\bigr)\,\max\,
|\,\delta g\,|.$$
Here $L[\,g(t),\,\delta g\,]$ is a functional linear with respect to $\delta g$, and $\max\,|\,\delta g\,|$ is the maximum value of $|\,\delta g\,|,$ where 
$$\beta\,\bigl(g(t),\,\delta g\bigr) \to 0\quad (\max\,|\,\delta g\,| \to 0).$$

\begin{definition}
The functional $F[\,g(t)\,]$ is called a primitive functional for the functional $f[\,g(t)\,]$ over the interval $[\,t_0,t_*\,]$ if the following equality holds in all points of this interval:
$$\delta\,F[\,g(t)\,] = f[\,g(t)\,]\,\delta\,g(t).$$
\end{definition}

Evidently, for the investigation of extremal properties of functionals, variation plays the same role as differential for investigation of similar properties of functions.

\begin{definition}
Let the functional $F[\,g(t)\,]$ be a primitive functional for the functional $f[\,g(t)\,].$ Then the expression
$$F[\,g(t)\,] = \int\,f[\,g(t)\,]\,\delta\,g(t)$$
is called an indefinite variational integral with respect to the integrating function $g(t).$
\end{definition}

{\bf Remarks.} 1. A usual indefinite integral is defined with an accuracy up to an arbitrary additive integration constant $C.$ Therein lies its difference from the indefinite variational integral, for which $\delta C \not= 0.$

2. Apparently, for the indefinite variational integral to exist it is necessary that the integrating function $g(t)$ be smooth of $r^{th}$ order, $r \ge 1,$ and the functional $f[\,g(t)\,]$ be continuous for all corresponding values of the function $g(t).$

From the definitions above, the following main integral equality follows as a consequence:
$$\delta\,\int\,f[\,g(t)\,]\,\delta\,g(t) = f[\,g(t)\,]\,\delta\,g(t).\eqno(49)$$
From equality (49) it follows immediately that
$$\delta\,F[\,g(t_0)\,] = \delta\,F[\,g(t_*)\,] = 0,$$
because $\delta\,g(t_0) = \delta\,g(t_*) = 0,$ i.e. the primitive functional $F[\,g(t)\,]$ should be fixed at the initial and end moments of time.

Variational integral over the range $[\,t_0,t_*\,]$ is defined in a somewhat more intricate way. As the time $t$ is not varied, the variational integral will be understood in the sense of an integral of vector functions $f(t),\,g(t)$ which are measurable on $[\,t_0,t_*\,]$.

Let us divide the segment $[\,t_0,t_*\,]$ into $n$ parts:
$$t_0 < t_1 < t_2 <\ ...\ < t_n = t_*,$$
where $\delta\,g(t_0) = \delta\,g(t_*) = 0.$ Now consider a composed vector function (vector functional) $f(t) = f[\,g(t)\,] \in C^r[\,t_0,t_*\,],\,r \ge 1.$ Suppose that for $s_k \in [\,t_k,\,t_{k+1}),\,k = \overline{0,n - 1},$ the following equality holds:
$f(t) = f(s_k) = f[\,g(s_k)\,].$

Let $H^2[\,t_0,t_*\,],\,0 \le t_0 < t_* \le \infty,$ denote the collection of composed vector functions, for which
$$\int_{t_)}^{t_*}\,|\,f(t)\,|^2\,dt < \infty,\qquad \|\,f\,\|_{H^2} = \biggl(
\int_{t_0}^{t_*}\,|\,f(t)\,|^2\,dt\biggr)^{1/2},$$
where $|\,f(t)\,|$ is the modulus of vector $f(t).$ For such functionals $f[\,g(t)\,|$ from $H^2[\,t_0,t_*\,]$ we define the variational integral with respect to the integrating vector function
$g(t),\,t \in [\,t_0,t_*\,],$ by setting
$$\int_{t_0}^{t_*}\,f[\,g(t)\,]\,\delta\,g(t) = \lim_{\max\,|\,\delta g(t)\,|
\to 0}\ \sum_{k=0}^{n-1}\,f[\,g(s_k)\,]\,\delta\,g(s_k),$$
where $\delta\,g(s_k) = \bar{g}(s_k) - g(s_k)$ is the variation of the function $g(t)$ in point $s_k,\,\bar{g}(s_k)$, is an admissible function in point $s_k.$

Hence, according to the definition given above, a step function (functional) $f$ is compared with the value 
$$F = \int_{t_0}^{t_*}\,f[\,g(t)\,]\,\delta\,g(t),$$
and this mapping preserves the norm. This mapping, initially defined for step functions, can be extended with preservation of the norm to the closure which, for the set of step functions in $H^2[\,t_0,t_*\,]$, will coincide with $H^2[\, t_0,t_*\,]$ (see [23, 24]). As a result, each element $f \in H^2[\,t_0,t_*\,]$ is associated with the value $F,$ which is called a definite variational integral of the function (functional) $f[\,g(t)\,]$ over the interval $[\,t_0,t_*\,]$ and is denoted as $\int_{t_0}^{t_*}\,f[\,g(t)\,]\,\delta\,g(t).$

It is important to note that the concept of a definite variational integral does not fit the mold of the visual geometric interpretation on a plane. The point is that though the variational integral here is also defined via the limit of the sequence of integral sums, but the components of these integral sums $f[\,g(s_k)\,]\,\delta \,g(s_k)$ are taken for one (!) moment of time $s_k.$ The role of a kind of "shift", though not in time but in space, is played by the variation of the function $\delta\,g(t).$ However, there are no objective logical and mathematical reasons to "refuse" to construct such a class of integrals and to introduce it into practice.

Let us also introduce the concept of a variational integral with a variable upper limit. To that end, using known relations, consider the value $\int_{t_0}^t
\,f[\,g(s)\,]\,\delta\,g(s),$ where $t \in [\,t_0,t_*\,].$ We have
$$\delta\,\biggl(\int_{t_0}^t\,f[\,g(s)\,]\,\delta\,g(s)\biggr) = f[\,g(s)\,]
\,\delta\,g(s)\,\big|_{t_0}^t$$
$$=\ f[\,g(t)\,]\,\delta\,g(t) = \delta\,F[\,g(t)\,],\qquad \delta\,g(t_0) = 0.$$

Eliminating the variation in right- and left-hand sides of this equation, we define a variational integral with a variable upper limit according to the following rule:
$$\int_{t_0}^t\,f[\,g(s)\,]\,\delta\,g(s) = F[\,g(t)\,] - F[\,g(t_0)\,].$$
The above equation holds because $\delta\,F[\,g(t_0)\,] = 0.$

This implies, in particular, important corollaries-analogues for the variational integral.

1. The formula for relation between an indefinite variational integral and a variational integral with variable upper limit (Barrow theorem) of the form
$$\int\,f[\,g(t)\,]\,\delta\,g(t) = \int_{t_0}^t\,f[\,g(s)\,]\,\delta\,g(s) +
F[\,g(t_0)\,].$$

2. Newton--Leibniz formula for a definite variational integral:
$$\int_{t_0}^{t_*}\,f[\,g(t)\,]\,\delta\,g(t) = F[\,g(t)\,]\,\big|_{t_0}^{t_*} =
F[\,g(t_*)\,] - F[\,g(t_0)\,].$$

The structure of the integral $\int\,f[\,g\,]\,\delta g$ allows for writing down Hamilton's principle (47) using relation (49) in the standard form. In fact, we have
$$\delta S_{H_*} = \delta\,\int_{t_0}^{t_*}\,L_*\,dt = \delta\,\int_{t_0}^{t_*}
\,\biggl(T_* - \int\,\dot{M}\dot{R}\,\delta R\biggr)\,dt = 0,\eqno(50)$$
where
$$S_{H_*} = \int_{t_0}^{t_*}\,L_*\,dt,\qquad L_* = T_* - \int\,\dot{M}\dot{R}\,
\delta R.$$
Here, $S_{H_*}$ is the {\it total $($generalized$)$ Hamiltonian action}, $L_*$
is the {\it generalized Lagrange function}, and the following equality holds in relation (50):
$$\delta\,\int\,\dot{M}\dot{R}\,\delta R = \dot{M}\dot{R}\,\delta R.$$

Summarizing the above, we can formulate the resulting statement, which represents Hamilton's principle for the general hyperreactive motion.

\begin{theorem}
The actual motion of a hyperreactive system under given potential forces and ideal, holonomic, stationary constraints in the time interval $[\,t_0,t_*\,]$ corresponds to such motion, for which the total Hamiltonian action $S_{H_*}$ $(50)$ takes on a stationary $($minimum$)$ value.
\end{theorem}

{\bf Remarks.} 1. From theorem 3, we obtain, as a corollary, Lagrange equations of the second kind (which are familiar in terms of the form) for generalized coordinates $q_1,q_2,\,...,\,q_s:$
$${d\over{dt}}\,{{\partial L_*}\over{\partial\dot{q}_j}} - {{\partial L_*}\over
{\partial q_j}} = 0,\qquad j = \overline{1,s},\eqno(51)$$
with respect to the generalized Lagrange function $L_*:$
$$L_* = {{M\dot{R}^2}\over 2} + \dot{M}\dot{R}R + U - \int\,\dot{M}\dot{R}\,
\delta R,$$
where the product of vectors is understood as their scalar product.

2. Comparing Lagrange equations (51) with respect to $L_*$ and (42) with respect to $T_*,$ one can conclude that equation (51) represents the most compact notation of the hyperreactive motion equation in generalized curvilinear coordinates.

We will complete this section which addresses the new type of variational integrals with a story about a new type of variations of functions that can be successfully used in general non-linear mechanics, in particular, for canonical transformations of variables. 

The matter is that there exists a certain class of so called {\it dynamic variations of generalized coordinates} $q_j,\,j = \overline{1,s},$ for which the equality of values of a variational and a conventional definite integrals is achieved. 

Let us introduce an important concept of the dynamical variation for function $q_j(t) \in C^r[\,t_0,t_*\,]$ in order to ensure the equality of variation and increment of the function $q_j(t):$
$$\delta\,q_j(s_k) = \Delta q_j(s_k,\,s_{k+1}),\qquad s_k \in [\,t_k,\,t_{k+1}),
\eqno(52)$$
where
$$\delta\,q_j(s_k) = \bar{q}_j(s_k) - q_j(s_k),\quad \Delta q_j(s_k,\,s_{k+1}) =
q_j(s_{k+1}) - q_j(s_k),$$
where $k = \overline{0,n - 1}.$ Evidently, equality (6.52) will hold if the following conditions are satisfied:
$$\bar{q}_j(s_k) = q_j(s_{k+1}),\qquad \bar{q}_j(s_0) = q_j(s_0),\qquad \bar{q}_j
(s_n) = q_j(s_n),\eqno(53)$$
where $s_0 = t_o,\,s_n = t_*.$

The latter relations are natural, as the function $\bar{q}_j(t)$ represents an "admissible" or "possible" function, values of which are close enough to the function $q_j(t).$ In equality (53), this admissibility of the function $\bar{q}_j(t)$ takes on the features of an actual "admissibility": in the next moment of time, the admissible function becomes an actually realized function. This means that relations (53) describe a set of physically and actually possible, that is, admissible functions which are described in time by specific differential equations for specific dynamic systems.

If relations (52) and (53) are fulfilled for the function $g(t)$, then the values of variational and conventional definite integrals are equal in the time interval $[\,t_0,t_*\,]$:
$$\int_{t_0}^{t_*}\,f[\,g(t)\,]\,\delta\,g(t) = \int_{t_0}^{t_*}\,f[\,g(t)\,]\,
d\,g(t),\eqno(54)$$
because both integrals in equality (54) represent the same limit of the sequence of integral sums
$$\lim_{\max\,|\,\delta g\,|\to 0}\,\sum_{k=0}^{n-1}\,f[\,g(s_k)\,]\,
\delta\,g(s_k) = \lim_{\max\,|\,\Delta g\,|\to 0}
\,\sum_{k=0}^{n-1}\,f[\,g(s_k)\,]\,\Delta g(s_k,\,s_{k+1}).$$

Hence, we can conclude that, taking into account that the {\it dynamic variation} condition (condition of physical realizability) is adopted for the reactive vector $R(t),$ for functions of generalized coordinates $q_j(t),\,j = \overline{1,s},$ and for their trajectories, Hamilton's principle in the hyperreactive case is formulated in the form of theorem 3. In this context, one should use relations (50), (51) with the generalized Lagrange function of the form
$$L_* = {{M\dot{R}^2}\over 2} + M\dot{R}R + U - \int\,\dot{M}\dot{R}\,dR,$$
where the indefinite integral $\int\,\dot{M}\dot{R}\,dR$ can be substituted with the integral with a variable upper limit $\int_{t_0}^t\,\dot{M}\dot{R}\,dR,$ if the following condition is satisfied: $\dot{M}\dot{R}\,dR\,\big|_{t=t_0}\,= 0.$

We complete this article with an indication of the author's works [25, 26] of the English version, devoted to the development of the theory of hyperreactive motion. Items [1-3, 5-7, 9-24] are published in Russian.

\section*{Conclusions}
In this article, which is of exploratory nature, the results and algorithmic formulae are obtained that allow for describing, in the most complete and accurate form, the general motion of a non-stationary hyperreactive dynamic system. Here is solved the sixth Hilbert's problem of mathematical axiomatization of dynamics taking into account the proposed axiom (principle) of completeness intended to generalize Newton's second law of motion for case of a non-stationary variable-mass point (system) that varies with time.

\end{document}